\documentclass[showpacs,
floatfix,
aps,
prb,
amsmath,
twocolumn,
superscriptaddress,
tightenlines]{revtex4-1}

\usepackage{graphicx}
\usepackage{dcolumn}
\usepackage{amsmath}
\usepackage{amsfonts}
\usepackage{bm}
\usepackage{color}

\newcommand{\be}{\begin{equation}}
\newcommand{\ee}{\end{equation}}
\newcommand{\bea}{\begin{eqnarray}}
\newcommand{\eea}{\end{eqnarray}}

\def\pc{\mathcal{P}}

\def\F{\mathcal{F}}

\def\tst{\tau_\star}
\def\eac{\epsilon}

\def\oc{\omega_{\mbox{\scriptsize {c}}}}

\def\tq{\tau_{\mbox{\scriptsize {q}}}}

\def\ttr{\tau}

\def\tin{\tau_{\mbox{\scriptsize {in}}}}

\newcommand{\req}[1]{Eq.\,(\ref{#1})}
\newcommand{\rEq}[1]{Equation (\ref{#1})}

\newcommand{\rfig}[1]{Fig.\,\ref{#1}}
\newcommand{\rFig}[1]{Figure \ref{#1}}

\begin{document}
\title{Observation of microwave-induced resistance oscillations in a high-mobility two-dimensional hole gas in Ge/SiGe quantum well}

\author{M.~A.~Zudov}
\affiliation{School of Physics and Astronomy, University of Minnesota, Minneapolis, Minnesota 55455, USA}
\author{O.~A.~Mironov}
\affiliation{Department of Physics, University of Warwick, Coventry, CV4 7AL, United Kingdom}
\affiliation{International Laboratory of High Magnetic Fields and Low Temperatures, 53-421 Wroclaw, Poland}
\author{Q.~A.~Ebner}
\author{P.~D.~Martin}
\author{Q.~Shi}
\affiliation{School of Physics and Astronomy, University of Minnesota, Minneapolis, Minnesota 55455, USA}
\author{D.~R.~Leadley}
\affiliation{Department of Physics, University of Warwick, Coventry, CV4 7AL, United Kingdom}

\begin{abstract}
Microwave-induced resistance oscillations (MIRO) have been extensively studied for more than a decade but, until now, have remained unique to GaAs/AlGaAs-based 2D electron systems.
Here, we report on the first observation of MIRO in a 2D hole gas hosted in Ge/SiGe quantum well.
Our findings confirm that MIRO is a universal phenomenon and demonstrate that microwave photoresistance can be utilized to probe the energy spectrum and the correlation effects of 2D holes in Ge/SiGe quantum wells.
\end{abstract}

\received{October 31, 2013}


\maketitle

%\textit{Introduction}.-- 
Two-dimensional electron gases (2DEG) subject to both perpendicular magnetic field $B$ and microwave radiation reveal a variety of fascinating non-equilibrium transport phenomena.\citep{dmitriev:2012}
Two prime examples of these phenomena are microwave-induced resistance oscillations (MIRO)\citep{zudov:2001a,ye:2001} and zero-resistance states,\citep{mani:2002,zudov:2003,yang:2003,andreev:2003,smet:2005,zudov:2006b,dorozhkin:2011} which emerge from the MIRO minima in ultra-high mobility 2DEG.\citep{note:11}
MIRO appear in microwave photoresistance, $\delta R_\omega$, which is the difference between the resistance of irradiated 2DEG and the resistance measured without radiation, $R_0$.
Theoretically, MIRO can be explained in terms of (i) the \emph{displacement} mechanism,\citep{ryzhii:1970,durst:2003,lei:2003,vavilov:2004,dmitriev:2009b} which is based on the shift of the cyclotron orbit due to microwave-assisted impurity scattering, and (ii) the \emph{inelastic} mechanism,\citep{dmitriev:2005,dmitriev:2009b} stemming from the radiation-induced modification of electron distribution function.
Both mechanisms predict that the photoresistance oscillates as 
\be
\delta R_\omega/R_0 = - 2\pi \eta \pc \lambda^2 \eac \sin 2\pi\eac\,,
\label{eq.miro}
\ee
where $\eta$ is the dimensionless scattering rate, which contains both displacement and inelastic contributions,\citep{dmitriev:2009b} $\pc$ is the dimensionless microwave power,\citep{dmitriev:2005,khodas:2008} $\lambda = \exp(-\pi/\oc\tq)$ is the Dingle factor, $\tq$ is the quantum lifetime, $\eac =\omega/\oc$, $\omega = 2\pi f$ is the microwave frequency, and $\oc=eB/m^\star$ is the cyclotron frequency of the charge carrier with the effective mass $m^\star$.

While MIRO have been actively investigated for more than a decade,\citep{dmitriev:2012} their observation has remained unique to \emph{n}-type GaAs/AlGaAs.
Indeed, experiments on microwave photoresistance in \emph{p}-type GaAs/AlGaAs,\citep{du:2004a} in \emph{n}-type Si/SiGe,\citep{sassine:2007} and in HgTe/CdHgTe\citep{kozlov:2011a} revealed only a single photoresistance peak owing to magneto-plasmon resonance.\citep{vasiliadou:1993}
Interestingly, microwave-induced magneto-conductance oscillations and associated zero-conductance states\citep{yang:2003} have been recently realized in a non-degenerate 2D system, electrons on liquid helium surface.\citep{konstantinov:2009,konstantinov:2010}
These oscillations, however, appear only under resonant excitation of the second subband and are very different from MIRO in many other aspects.\citep{monarkha:2011}

%extended abstract
In this paper we report on a first observation of MIRO in a new material system, a high-mobility 2D hole gas (2DHG) hosted in a pure Ge/SiGe quantum well.
First, we have found that MIRO exist over a wide range of microwave frequencies and are well described by the hole effective mass of $m^\star \approx 0.09 m_0$, where $m_0$ is a mass of a free electron.
Second, we have observed that with increasing temperature $T$, the MIRO amplitude decays roughly as $T^{-2}$, suggesting the dominance of the inelastic mechanism.\citep{dmitriev:2005}
Finally, we have found that MIRO exhibit a sublinear dependence on microwave power, consistent with many MIRO experiments in GaAs/AlGaAs.\citep{ye:2001,studenikin:2004,mani:2004a,hatke:2011e}
Taken together, these findings establish that MIRO are not unique to GaAs/AlGaAs and demonstrate that microwave photoresistance can be used to probe the energy spectrum and correlations of 2D holes in Ge/SiGe quantum wells.
Future experiments utilizing higher microwave frequencies and dc electric fields should yield unique information on the correlation properties of the disorder potential.

%sample and experimental details
Our sample was fabricated from a fully strained (0.65\,\%), 20-nm wide, 99.99\,\% pure (Si-free) Ge quantum well grown on Si$_{0.2}$Ge$_{0.8}$ by reduced pressure chemical vapor deposition.\citep{dobbie:2012,note:10}
Holes were supplied by a 10 nm-wide B-doped layer separated from Ge channel by a 26 nm-wide undoped Si$_{0.2}$Ge$_{0.8}$ spacer. 
The sample was a 4$\times$4 mm square with ohmic contacts formed by thermal evaporation of Al followed by annealing in N$_2$ at $425~^\circ$C.
At $T = 1.5$ K, our 2DHG has the hole density $p \approx 2.8\cdot 10 ^{11}$ cm$^{-2}$ and the mobility $\mu \approx 0.4\cdot 10^{6}$ cm$^2$/V$\cdot$s.\citep{note:6,dobbie:2012,hassan:2013}
Resistance measurements were performed at $T$ from 1 K to 5 K using a low-frequency (13 Hz) lock-in detection in sweeping $B$.
Radiation ($f = 30-110$ GHz) was delivered to the sample via a rectangular (WR-28) waveguide. 

%%%%%%%%%%%%%%%%%%%%%%%%%%%%%%%%%%%%%%%%%%%%%%%%%%%
\begin{figure}[t]
\includegraphics{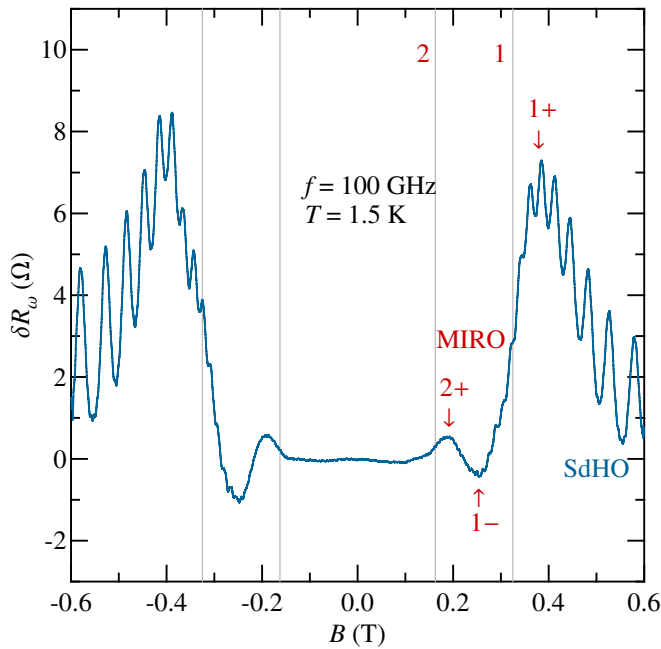}
\vspace{-0.05 in}
\caption{
Microwave photoresistance $\delta R_\omega$ as function of $B$ measured at $f = 100$ GHz and $T=$ 1.5 K.
Vertical lines are drawn at $\epsilon = 1$ and 2, as marked, computed using $m^* = 0.091 m_0$. 
}
\vspace{-0.15 in}
\label{trace}
\end{figure}

A typical example of microwave photoresistance $\delta R_\omega$ measured at $f = 100$ GHz and $T = 1.5$ K is shown in \rfig{trace} as a function of magnetic field.
At $B\gtrsim 0.3$ T the data reveal fast oscillations which reflect a reduction of the amplitude of Shubnikov-de Haas oscillations in irradiated 2DHG due to radiation-induced heating.
In addition, one observes another oscillatory structure which persists down to considerably lower magnetic fields, $B \approx 0.13$ T.
This structure is represented by two photoresistance maxima (marked by $1+,2+$) and one minimum (marked by $1-$).
One can clearly see that the photoresistance at the minimum is \emph{negative} indicating that microwave radiation causes a \emph{reduction} of resistance from its dark value.
This negative $\delta R_\omega$ is one characteristic feature of MIRO which, in very clean GaAs/AlGaAs-based 2DEG, gives rise to radiation-induced zero-resistance states.\citep{mani:2002,zudov:2003,yang:2003,andreev:2003,smet:2005,zudov:2006b,dorozhkin:2011}

Another generic MIRO feature is related to the positions of the maxima and minima relative to the harmonics of the cyclotron resonance.
According to \req{eq.miro}, MIRO maxima (minima) occur near $\eac_{i+}$ ($\eac_{i-}$) given by
\be
\eac_{i\pm} = i \mp 1/4\,,~i = 1,2,3,...\,.
\label{eq.m}
\ee
\rEq{eq.m} thus prescribes that MIRO maxima and minima should appear roughly symmetrically offset by a quarter-cycle from the harmonics of the cyclotron resonance, $\eac = i$.
As illustrated by vertical lines (marked by 1, 2) in \rfig{trace}, measured photoresistance $\delta R_\omega$ conforms to \req{eq.m} reasonably well if one uses the effective hole mass value of $m^\star = 0.091\,m_0$ to calculate the harmonics of the cyclotron resonance, $\eac = i = 1,2$.

%FIGURE 2
%%%%%%%%%%%%%%%%%%%%%%%%%%%
\begin{figure}[t]
\includegraphics{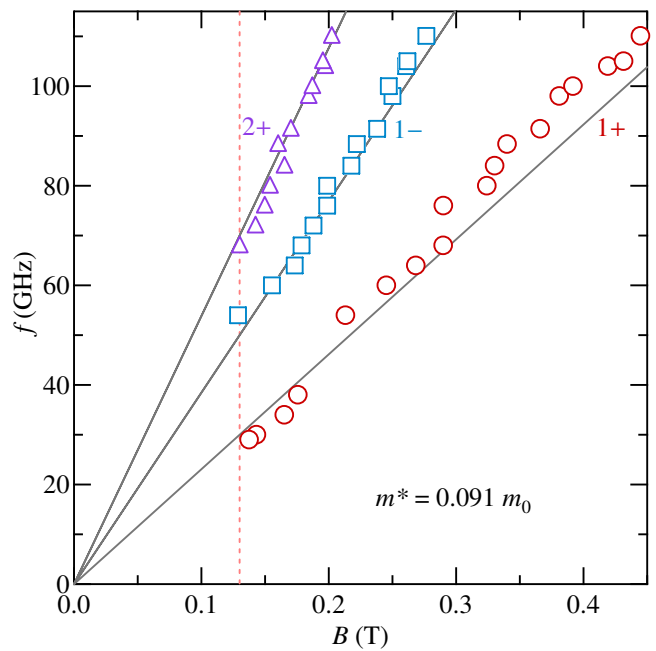}
\vspace{-0.15 in}
\caption{[Color online] Frequency $f$ as a function of $B$ corresponding to the first MIRO maximum ($1+$, circles), the first minimum ($1-$, squares), and the second maximum ($2+$, triangles). 
Solid lines correspond to $\epsilon_{1+}$, $\epsilon_{1-}$, and $\epsilon_{2+}$, computed using \req{eq.m} with $m^\star = 0.091 m_0$.
Vertical line marks the lowest $B$, $B_0 = 0.13$ T, at which MIRO first appear.
}
\vspace{-0.1 in}
\label{fan}
\end{figure}
%%%%%%%%%%%%%%%%%%%%%%%%%%%%
\rEq{eq.m} also predicts that the $B$ positions of the MIRO maxima and minima should scale \emph{linearly} with $f$, according to $f_{i\pm} = e B_{i\pm} (i \mp 1/4)/2\pi m^\star$.
To verify this prediction we have carried out measurements at a variety of microwave frequencies, from 30 to 110 GHz, at $T = 1.5$ K.
The results of this study are summarized in \rfig{fan} showing $f$ as a function of $B$ at the first maximum ($1+$, circles), the first minimum ($1-$, squares), and the second maximum ($2+$, triangles) of $\delta R_\omega$. 
Solid lines emanating from the origin correspond to $\epsilon_{1+}$, $\epsilon_{1-}$, and $\epsilon_{2+}$, computed using \req{eq.m} with $m^\star = 0.091 m_0$.
We observe good overall agreement between the experimental data and \req{eq.m} over the entire frequency range.
This finding confirms that $\delta R_\omega$ in our 2DHG is periodic in $\eac$, with a period equal to unity, in agreement with \req{eq.miro}.
We therefore conclude that the photoresistance of 2DHG in Ge/SiGe possesses all characteristic MIRO properties.

Another interesting feature of the data presented in \rfig{fan} is that all MIRO extrema, independent of their order, cease to exist below a certain, well-defined magnetic field.
As illustrated by a vertical dotted line, drawn at $B_0 = 0.13$ T, this observation holds true for all frequencies studied.
The value of $B_0$ can therefore be used to obtain a rough estimate of quantum lifetime in our 2DHG; setting $\oc\tq = 1$ at $B=B_0$ yields $\tq = m^\star/eB_0 \approx$ 4 ps.
% \citep{note:0}
This value is considerably lower than $\tq$ usually found in ultra-high mobility 2DEG, where it typically ranges from 10 to 20 ps,\citep{zhang:2007a,zhang:2007c,hatke:2009a,hatke:2011b,bogan:2012} explaining why only two MIRO maxima are detected in our data.
It would be interesting to extend the experiments to higher radiation frequencies, where one expects to see more oscillations, and accurately determine $\tq$ using a standard Dingle plot procedure.\citep{hatke:2009a}

%FIGURE 3
%%%%%%%%%%%%%%%%%%%%%%%%%%%%%%
\begin{figure}[t]
\includegraphics{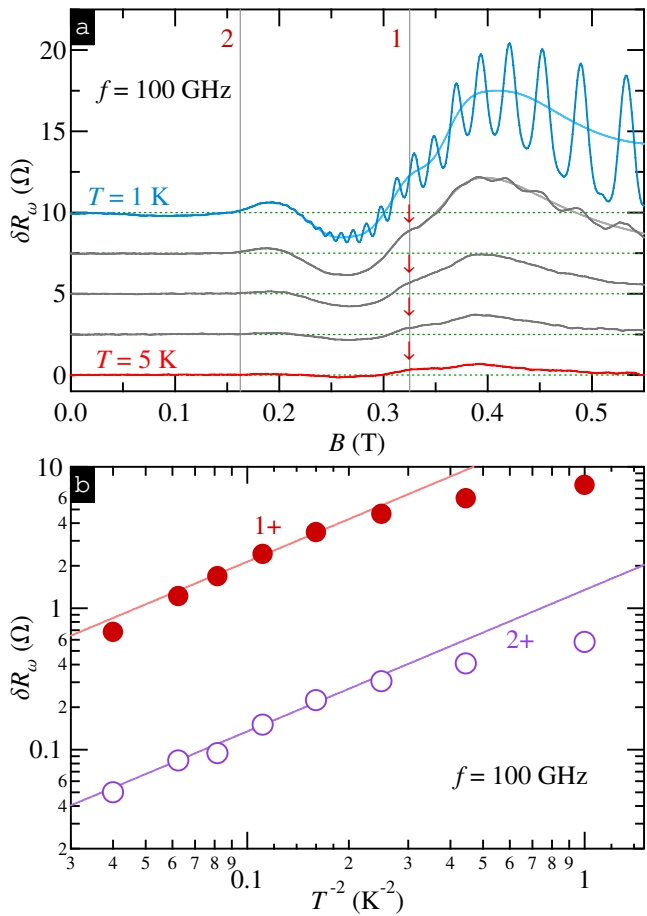}
\vspace{-0.1 in}
\caption{[Color online] (a) Photoresistance $\delta R_\omega$ as a function of $B$ measured at $f = 100$ GHz and $T$ from 1 K to 5 K, in steps of 1 K. 
Smooth curves represent the photoresistance without Shubnikov-de Haas component. 
The traces are vertically offset for clarity by $2.5$ $\Omega$.
Vertical lines are drawn at $\epsilon = 1,2$, as marked. 
Arrows mark an extra feature which emerges near the cyclotron resonance (cf. $\downarrow$).
(b) $\delta R_\omega$ at $\eac=\eac_{1+}$ (solid circles) and at $\eac=\eac_{2+}$ (open circles) as a function of $T^{-2}$ on a log-log scale.
Solid lines mark $\delta R_\omega \propto T^{-2}$ dependence.
}
\label{tmp}
\vspace{-0.15 in}
\end{figure}
%%%%%%%%%%%%%%%%%%%%%%%%%%%%%%%%%%

We next discuss the temperature dependence of MIRO in our 2DHG.
In \rfig{tmp}(a) we present $\delta R_\omega$ as a function of $B$ measured at $f = 100$ GHz and different $T$, from 1 K to 5 K, in steps of 1 K. 
The traces are vertically offset for clarity by $2.5$ $\Omega$ and vertical lines are drawn at $\epsilon = 1,2$, as marked. 
As anticipated, we observe that MIRO gradually decay with increasing $T$. 

In general, the decay of MIRO can be due to two distinct factors.
One of these factors stems from the carrier-carrier interaction-induced correction to the quantum scattering rate, $\tq^{-1}$,\citep{hatke:2009a,hatke:2009b,hatke:2009c} which is of the order of $\tin^{-1} \sim T^2/E_F$ ($E_F$ is the Fermi energy)\citep{giuliani:1982,dmitriev:2003,dmitriev:2005} and enters $\lambda^2 =\exp(-2\pi/\oc\tq)$ appearing in \req{eq.miro}.  
This correction can be quite significant in ultra high-mobility 2DEG, where it can approach and even exceed the disorder contribution at $T$ of a few kelvin.\citep{hatke:2009a,hatke:2009b,hatke:2009c}
In our 2DHG, however, the impurity scattering rate is considerably larger and, as a result, one cannot expect significant interaction-induced increase of $\tq^{-1}$.
Indeed, at $T = 4$ K, we estimate $\tin^{-1} \approx 0.2$ K, which is an order of magnitude smaller than impurity contribution $\tq^{-1} \approx 2$ K.\citep{note:8}

Another $T$-dependent factor appears in the dimensionless scattering rate $\eta$, entering \req{eq.miro}, owing to the inelastic contribution.
More specifically,
\be
\eta = \frac{\ttr}{2\tst} + \frac {2\tin}{\ttr}\,,
\label{eta}
\ee
where the first (second) term represents displacement\citep{dmitriev:2009b} (inelastic\citep{dmitriev:2005}) contribution.
We first recall that $\ttr/2\tst$ is determined by the correlation properties of the disorder potential.\citep{note:4} 
For purely smooth disorder (e.g. from remote ionized acceptors), one finds $\ttr/2\tst  \sim 6 \tq/\ttr$.\citep{dmitriev:2005}
In the opposite limit of only sharp disorder (e.g. from residual impurities in the quantum well), the factor representing displacement contribution attains its maximal possible value, $\ttr/2\tst = 3/2$.
As a result, regardless of the specifics of the disorder potential, $\ttr/2\tst \lesssim 1$.

The inelastic contribution, given by ${2\tin}/{\ttr}$, is controlled by the inelastic relaxation time, $\tin \sim E_F/T^2$, and therefore can be dominant at low $T$, especially in lower mobility samples.
At $T = 4$ K, we estimate $\tin \simeq 40$ ps, $2\tin/\ttr \approx 4$, and conclude that the inelastic contribution should dominate the microwave photoresistance over the entire temperature range.

Based on the above picture, the temperature dependence of the MIRO amplitude can be described by a function $\F(T) \simeq 1+T_\star^2/T^2 \approx T_\star^2/T^2$, where $T_\star \sim 10$ K.
We thus expect that the MIRO amplitude should decay roughly as $T^{-2}$ with increasing temperature.
To check this prediction we plot the MIRO amplitude $\delta R_\omega$ at $\eac = \eac_{1+}$ (solid circles) and at $\eac = \eac_{2+}$ (open circles) as a function of $T^{-2}$ in \rfig{tmp}(a) on a log-log scale.
At higher temperatures, we observe good agreement with $T^{-2}$-dependence, which is illustrated by solid line.\citep{note:1}
At lower temperatures, however, the data fall below the expected dependence which can be attributed to the radiation-induced heating of the 2DHG.
The observed temeprature dependence of the oscillation amplitude suggests that MIRO originate from the inelastic mechanism of microwave photoresistance.\citep{dmitriev:2005}

More careful examination of the data in \rfig{tmp}(a) also reveals an extra photoresistance feature which occurs close to the cyclotron resonance (cf.\,$\downarrow$). 
While the exact origin of this feature is unclear at this point, its shape and position are consistent\citep{note:2} with the dimensional magnetoplasmon resonance which is frequently detected in experiments using GaAs/AlGaAs quantum wells.\citep{vasiliadou:1993,zudov:2001a,hatke:2012b,hatke:2013}
To confirm the origin of this feature, future experiments should employ Hall bar-shaped 2DHG, where the magnetoplasmon dispersion differs significantly from that of the cyclotron resonance.

%FIGURE 4
%%%%%%%%%%%%%%%%%%%%%%%%%%%%%%
\begin{figure}[t]
\includegraphics{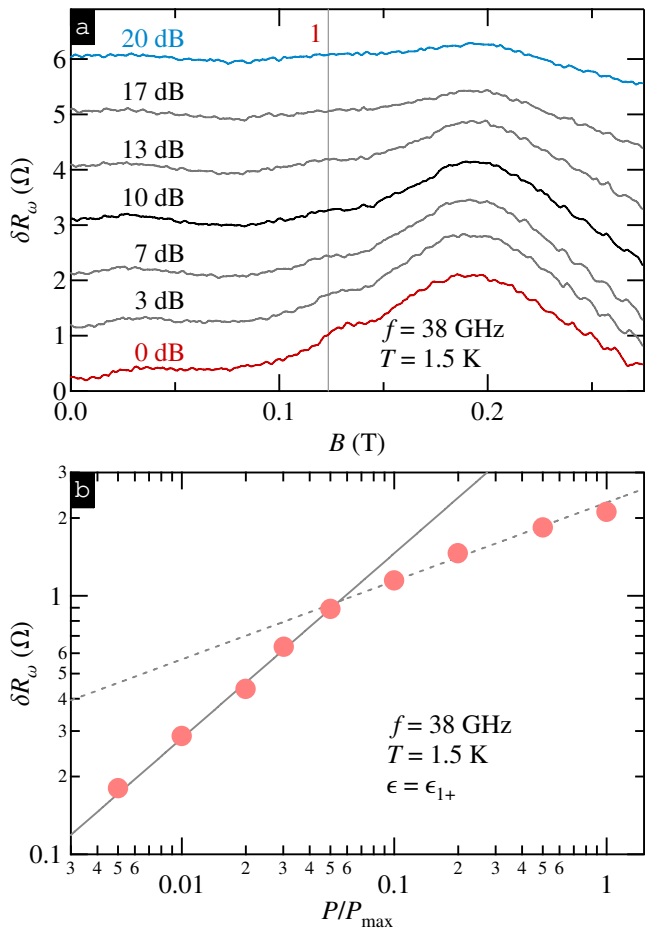}
\vspace{-0.1 in}
\caption{[Color online] (a) Photoresistance $\delta R_\omega$ as a function of $B$ measured at $f = 38$ GHz, $T = 1.5$ K, and different microwave powers, marked by attenuation factors, from 0 dB to 20 dB.
The traces are vertically offset for clarity by $1$ $\Omega$.
(b) $\delta R_\omega$ at $\eac=\eac_{1+}$ a function of $P$ on a log-log scale.
Solid (dotted) line marks $\delta R_\omega \propto P^{0.7}$ ($\delta R_\omega \propto P^{0.3}$) dependence.
}
\label{pow}
\vspace{-0.15 in}
\end{figure}
%%%%%%%%%%%%%%%%%%%%%%%%%%%%%%%%%%
%POWER DEPENDENCE
Finally, we briefly discuss the power dependence of MIRO in our 2DHG. 
\rFig{pow}(a) shows $\delta R_\omega$ as a function of $B$ measured at $f = 38$ GHz,\citep{note:7} $T = 1.5$ K, and different microwave powers, marked by attenuation factors, from 0 dB to 20 dB.
The traces are vertically offset for clarity by 1 $\Omega$.
At this microwave frequency we are able to detect only the fundamental MIRO maximum which occurs at $B \approx 0.17$ T and becomes weaker with decreasing power.
In \rfig{pow}(b) we present the magnitude of this MIRO peak as a function of microwave power $P$ (normalized to its maximum value) on a log-log scale.
At lower intensities, the amplitude grows roughly as $P^\alpha$, with $\alpha \approx 0.7$ (cf.\,solid line).
At higher intensities, the dependence gets weaker and the data are best described by $P^\alpha$ with $\alpha \approx 0.3$ (cf.\,dotted line).
Theoretically, one expects that the linear $P$ dependence, $\delta R_\omega \propto P$ crosses over to the square-root dependence, $\delta R_\omega \propto \sqrt{P}$, with increasing $P$, as a result of saturation effects\citep{dmitriev:2005,hatke:2011e} or increased importance of multiphoton processes.\citep{khodas:2008,khodas:2010,hatke:2011e}
While this crossover has been recently observed in GaAs/AlGaAs-based 2DEG,\citep{hatke:2011e} there exist a number of experiments reporting only sublinear power dependence with a rather wide range of exponents.\citep{ye:2001,studenikin:2004,mani:2004a,mani:2010}

%Summary
In summary, we have observed microwave-induced resistance oscillations in 2DHG hosted in a pure Ge/SiGe quantum well, demonstrating that MIRO are not restricted to 2DEG in GaAs/AlGaAs.
We have found that MIRO are well described by the hole effective mass of $m^\star \approx 0.09 m_0$ over the wide range of microwave frequencies.
We have further shown that the MIRO amplitude decays as $T^{-2}$ indicating the dominant contribution of the inelastic mechanism.\citep{dmitriev:2005}
Finally, we have observed that MIRO exhibits a sublinear dependence on microwave power, consistent with many MIRO experiments on GaAs/AlGaAs.

Observation of MIRO in Ge/SiGe opens a pathway to further interesting experiments.
In particular, employing higher microwave frequencies should increase the number of observed oscillations allowing a reliable measurement of quantum lifetime, knowledge of which is crucial for understanding the disorder potential in Ge/SiGe systems.\citep{note:9}
It will also be interesting to see if other nonlinear phenomena, such as Hall field-induced resistance oscillations,\citep{yang:2002,zhang:2007a,zhang:2007c} can be realized in Ge/SiGe-based 2DHG.
Such observation would provide direct information on the amount of background impurities in the Ge channel which cannot be obtained from  conventional transport methods.
Understanding transport properties of this material system might be important for future Ge-based devices which are attractive candidates for non-silicon-based semiconductor technology.\citep{pillarisetty:2011}

We thank R. A. Roemer for critical reading of the manuscript, A. Dobbie for Ge/Si$_{0.2}$Ge$_{0.8}$ wafer growth, A. H. A. Hassan for device fabrication, and R. J. H. Morris for SIMS/XRD characterization and discussions.
The work at University of Minnesota was funded by the US Department of Energy, Office of Basic Energy Sciences, under Grant No. DE-SC002567. 
The work at Warwick University was partially supported by the EPSRC UK Renaissance Germanium project.
O.A.M. acknowledges the support of ILHMFLT (Wroclaw, Poland), SAS Centre of Excellence CFNT MVEP (Kosice, Slovakia),  DFG project DR832/3-1 at HLD-HZDR (Rossendorf, Germany) as a member of the European Magnetic Field Laboratory (EMFL), and National Scholarship Program of the Slovak Republic for the Mobility Support for Researchers in the Academic Year 2012/2013.

%\bibliographystyle{../../apsrev}
%\bibliography{../../bibRMP1,footnotes}

\begin{thebibliography}{54}
\expandafter\ifx\csname natexlab\endcsname\relax\def\natexlab#1{#1}\fi
\expandafter\ifx\csname bibnamefont\endcsname\relax
  \def\bibnamefont#1{#1}\fi
\expandafter\ifx\csname bibfnamefont\endcsname\relax
  \def\bibfnamefont#1{#1}\fi
\expandafter\ifx\csname citenamefont\endcsname\relax
  \def\citenamefont#1{#1}\fi
\expandafter\ifx\csname url\endcsname\relax
  \def\url#1{\texttt{#1}}\fi
\expandafter\ifx\csname urlprefix\endcsname\relax\def\urlprefix{URL }\fi
\providecommand{\bibinfo}[2]{#2}
\providecommand{\eprint}[2][]{\url{#2}}

\bibitem[{\citenamefont{Dmitriev et~al.}(2012)\citenamefont{Dmitriev, Mirlin,
  Polyakov, and Zudov}}]{dmitriev:2012}
\bibinfo{author}{\bibfnamefont{I.~A.} \bibnamefont{Dmitriev}},
  \bibinfo{author}{\bibfnamefont{A.~D.} \bibnamefont{Mirlin}},
  \bibinfo{author}{\bibfnamefont{D.~G.} \bibnamefont{Polyakov}},
  \bibnamefont{and} \bibinfo{author}{\bibfnamefont{M.~A.} \bibnamefont{Zudov}},
  \bibinfo{journal}{Rev. Mod. Phys.} \textbf{\bibinfo{volume}{84}},
  \bibinfo{pages}{1709} (\bibinfo{year}{2012}).

\bibitem[{\citenamefont{Zudov et~al.}(2001)\citenamefont{Zudov, Du, Simmons,
  and Reno}}]{zudov:2001a}
\bibinfo{author}{\bibfnamefont{M.~A.} \bibnamefont{Zudov}},
  \bibinfo{author}{\bibfnamefont{R.~R.} \bibnamefont{Du}},
  \bibinfo{author}{\bibfnamefont{J.~A.} \bibnamefont{Simmons}},
  \bibnamefont{and} \bibinfo{author}{\bibfnamefont{J.~L.} \bibnamefont{Reno}},
  \bibinfo{journal}{Phys. Rev. B} \textbf{\bibinfo{volume}{64}},
  \bibinfo{pages}{201311(R)} (\bibinfo{year}{2001}).

\bibitem[{\citenamefont{Ye et~al.}(2001)\citenamefont{Ye, Engel, Tsui, Simmons,
  Wendt, Vawter, and Reno}}]{ye:2001}
\bibinfo{author}{\bibfnamefont{P.~D.} \bibnamefont{Ye}},
  \bibinfo{author}{\bibfnamefont{L.~W.} \bibnamefont{Engel}},
  \bibinfo{author}{\bibfnamefont{D.~C.} \bibnamefont{Tsui}},
  \bibinfo{author}{\bibfnamefont{J.~A.} \bibnamefont{Simmons}},
  \bibinfo{author}{\bibfnamefont{J.~R.} \bibnamefont{Wendt}},
  \bibinfo{author}{\bibfnamefont{G.~A.} \bibnamefont{Vawter}},
  \bibnamefont{and} \bibinfo{author}{\bibfnamefont{J.~L.} \bibnamefont{Reno}},
  \bibinfo{journal}{Appl. Phys. Lett.} \textbf{\bibinfo{volume}{79}},
  \bibinfo{pages}{2193} (\bibinfo{year}{2001}).

\bibitem[{\citenamefont{Mani et~al.}(2002)\citenamefont{Mani, Smet, von
  Klitzing, Narayanamurti, Johnson, and Umansky}}]{mani:2002}
\bibinfo{author}{\bibfnamefont{R.~G.} \bibnamefont{Mani}},
  \bibinfo{author}{\bibfnamefont{J.~H.} \bibnamefont{Smet}},
  \bibinfo{author}{\bibfnamefont{K.}~\bibnamefont{von Klitzing}},
  \bibinfo{author}{\bibfnamefont{V.}~\bibnamefont{Narayanamurti}},
  \bibinfo{author}{\bibfnamefont{W.~B.} \bibnamefont{Johnson}},
  \bibnamefont{and} \bibinfo{author}{\bibfnamefont{V.}~\bibnamefont{Umansky}},
  \bibinfo{journal}{Nature (London)} \textbf{\bibinfo{volume}{420}},
  \bibinfo{pages}{646} (\bibinfo{year}{2002}).

\bibitem[{\citenamefont{Zudov et~al.}(2003)\citenamefont{Zudov, Du, Pfeiffer,
  and West}}]{zudov:2003}
\bibinfo{author}{\bibfnamefont{M.~A.} \bibnamefont{Zudov}},
  \bibinfo{author}{\bibfnamefont{R.~R.} \bibnamefont{Du}},
  \bibinfo{author}{\bibfnamefont{L.~N.} \bibnamefont{Pfeiffer}},
  \bibnamefont{and} \bibinfo{author}{\bibfnamefont{K.~W.} \bibnamefont{West}},
  \bibinfo{journal}{Phys. Rev. Lett.} \textbf{\bibinfo{volume}{90}},
  \bibinfo{pages}{046807} (\bibinfo{year}{2003}).

\bibitem[{\citenamefont{Yang et~al.}(2003)\citenamefont{Yang, Zudov, Knuuttila,
  Du, Pfeiffer, and West}}]{yang:2003}
\bibinfo{author}{\bibfnamefont{C.~L.} \bibnamefont{Yang}},
  \bibinfo{author}{\bibfnamefont{M.~A.} \bibnamefont{Zudov}},
  \bibinfo{author}{\bibfnamefont{T.~A.} \bibnamefont{Knuuttila}},
  \bibinfo{author}{\bibfnamefont{R.~R.} \bibnamefont{Du}},
  \bibinfo{author}{\bibfnamefont{L.~N.} \bibnamefont{Pfeiffer}},
  \bibnamefont{and} \bibinfo{author}{\bibfnamefont{K.~W.} \bibnamefont{West}},
  \bibinfo{journal}{Phys. Rev. Lett.} \textbf{\bibinfo{volume}{91}},
  \bibinfo{pages}{096803} (\bibinfo{year}{2003}).

\bibitem[{\citenamefont{Andreev et~al.}(2003)\citenamefont{Andreev, Aleiner,
  and Millis}}]{andreev:2003}
\bibinfo{author}{\bibfnamefont{A.~V.} \bibnamefont{Andreev}},
  \bibinfo{author}{\bibfnamefont{I.~L.} \bibnamefont{Aleiner}},
  \bibnamefont{and} \bibinfo{author}{\bibfnamefont{A.~J.}
  \bibnamefont{Millis}}, \bibinfo{journal}{Phys. Rev. Lett.}
  \textbf{\bibinfo{volume}{91}}, \bibinfo{pages}{056803}
  (\bibinfo{year}{2003}).

\bibitem[{\citenamefont{Smet et~al.}(2005)\citenamefont{Smet, Gorshunov, Jiang,
  Pfeiffer, West, Umansky, Dressel, Meisels, Kuchar, and von
  Klitzing}}]{smet:2005}
\bibinfo{author}{\bibfnamefont{J.~H.} \bibnamefont{Smet}},
  \bibinfo{author}{\bibfnamefont{B.}~\bibnamefont{Gorshunov}},
  \bibinfo{author}{\bibfnamefont{C.}~\bibnamefont{Jiang}},
  \bibinfo{author}{\bibfnamefont{L.}~\bibnamefont{Pfeiffer}},
  \bibinfo{author}{\bibfnamefont{K.}~\bibnamefont{West}},
  \bibinfo{author}{\bibfnamefont{V.}~\bibnamefont{Umansky}},
  \bibinfo{author}{\bibfnamefont{M.}~\bibnamefont{Dressel}},
  \bibinfo{author}{\bibfnamefont{R.}~\bibnamefont{Meisels}},
  \bibinfo{author}{\bibfnamefont{F.}~\bibnamefont{Kuchar}}, \bibnamefont{and}
  \bibinfo{author}{\bibfnamefont{K.}~\bibnamefont{von Klitzing}},
  \bibinfo{journal}{Phys. Rev. Lett.} \textbf{\bibinfo{volume}{95}},
  \bibinfo{pages}{116804} (\bibinfo{year}{2005}).

\bibitem[{\citenamefont{Zudov et~al.}(2006)\citenamefont{Zudov, Du, Pfeiffer,
  and West}}]{zudov:2006b}
\bibinfo{author}{\bibfnamefont{M.~A.} \bibnamefont{Zudov}},
  \bibinfo{author}{\bibfnamefont{R.~R.} \bibnamefont{Du}},
  \bibinfo{author}{\bibfnamefont{L.~N.} \bibnamefont{Pfeiffer}},
  \bibnamefont{and} \bibinfo{author}{\bibfnamefont{K.~W.} \bibnamefont{West}},
  \bibinfo{journal}{Phys. Rev. Lett.} \textbf{\bibinfo{volume}{96}},
  \bibinfo{pages}{236804} (\bibinfo{year}{2006}).

\bibitem[{\citenamefont{Dorozhkin et~al.}(2011)\citenamefont{Dorozhkin,
  Pfeiffer, West, von Klitzing, and Smet}}]{dorozhkin:2011}
\bibinfo{author}{\bibfnamefont{S.~I.} \bibnamefont{Dorozhkin}},
  \bibinfo{author}{\bibfnamefont{L.}~\bibnamefont{Pfeiffer}},
  \bibinfo{author}{\bibfnamefont{K.}~\bibnamefont{West}},
  \bibinfo{author}{\bibfnamefont{K.}~\bibnamefont{von Klitzing}},
  \bibnamefont{and} \bibinfo{author}{\bibfnamefont{J.~H.} \bibnamefont{Smet}},
  \bibinfo{journal}{Nature Phys.} \textbf{\bibinfo{volume}{7}},
  \bibinfo{pages}{336} (\bibinfo{year}{2011}).

\bibitem[{not({\natexlab{a}})}]{note:11}
\bibinfo{note}{ZRS have also been realized in high density ($\sim 10^{12}$
  cm$^{-2}$), moderate mobility ($\approx 5\times10^{5}$ cm$^2$/V$\cdot$s) 2DEG
  in GaAs/AlGaAs.\citep{bykov:2006}}.

\bibitem[{\citenamefont{Ryzhii}(1970)}]{ryzhii:1970}
\bibinfo{author}{\bibfnamefont{V.~I.} \bibnamefont{Ryzhii}},
  \bibinfo{journal}{Sov. Phys. Solid State} \textbf{\bibinfo{volume}{11}},
  \bibinfo{pages}{2078} (\bibinfo{year}{1970}).

\bibitem[{\citenamefont{Durst et~al.}(2003)\citenamefont{Durst, Sachdev, Read,
  and Girvin}}]{durst:2003}
\bibinfo{author}{\bibfnamefont{A.~C.} \bibnamefont{Durst}},
  \bibinfo{author}{\bibfnamefont{S.}~\bibnamefont{Sachdev}},
  \bibinfo{author}{\bibfnamefont{N.}~\bibnamefont{Read}}, \bibnamefont{and}
  \bibinfo{author}{\bibfnamefont{S.~M.} \bibnamefont{Girvin}},
  \bibinfo{journal}{Phys. Rev. Lett.} \textbf{\bibinfo{volume}{91}},
  \bibinfo{pages}{086803} (\bibinfo{year}{2003}).

\bibitem[{\citenamefont{Lei and Liu}(2003)}]{lei:2003}
\bibinfo{author}{\bibfnamefont{X.~L.} \bibnamefont{Lei}} \bibnamefont{and}
  \bibinfo{author}{\bibfnamefont{S.~Y.} \bibnamefont{Liu}},
  \bibinfo{journal}{Phys. Rev. Lett.} \textbf{\bibinfo{volume}{91}},
  \bibinfo{pages}{226805} (\bibinfo{year}{2003}).

\bibitem[{\citenamefont{Vavilov and Aleiner}(2004)}]{vavilov:2004}
\bibinfo{author}{\bibfnamefont{M.~G.} \bibnamefont{Vavilov}} \bibnamefont{and}
  \bibinfo{author}{\bibfnamefont{I.~L.} \bibnamefont{Aleiner}},
  \bibinfo{journal}{Phys. Rev. B} \textbf{\bibinfo{volume}{69}},
  \bibinfo{pages}{035303} (\bibinfo{year}{2004}).

\bibitem[{\citenamefont{Dmitriev et~al.}(2009)\citenamefont{Dmitriev, Khodas,
  Mirlin, Polyakov, and Vavilov}}]{dmitriev:2009b}
\bibinfo{author}{\bibfnamefont{I.~A.} \bibnamefont{Dmitriev}},
  \bibinfo{author}{\bibfnamefont{M.}~\bibnamefont{Khodas}},
  \bibinfo{author}{\bibfnamefont{A.~D.} \bibnamefont{Mirlin}},
  \bibinfo{author}{\bibfnamefont{D.~G.} \bibnamefont{Polyakov}},
  \bibnamefont{and} \bibinfo{author}{\bibfnamefont{M.~G.}
  \bibnamefont{Vavilov}}, \bibinfo{journal}{Phys. Rev. B}
  \textbf{\bibinfo{volume}{80}}, \bibinfo{pages}{165327}
  (\bibinfo{year}{2009}).

\bibitem[{\citenamefont{Dmitriev et~al.}(2005)\citenamefont{Dmitriev, Vavilov,
  Aleiner, Mirlin, and Polyakov}}]{dmitriev:2005}
\bibinfo{author}{\bibfnamefont{I.~A.} \bibnamefont{Dmitriev}},
  \bibinfo{author}{\bibfnamefont{M.~G.} \bibnamefont{Vavilov}},
  \bibinfo{author}{\bibfnamefont{I.~L.} \bibnamefont{Aleiner}},
  \bibinfo{author}{\bibfnamefont{A.~D.} \bibnamefont{Mirlin}},
  \bibnamefont{and} \bibinfo{author}{\bibfnamefont{D.~G.}
  \bibnamefont{Polyakov}}, \bibinfo{journal}{Phys. Rev. B}
  \textbf{\bibinfo{volume}{71}}, \bibinfo{pages}{115316}
  (\bibinfo{year}{2005}).

\bibitem[{\citenamefont{Khodas and Vavilov}(2008)}]{khodas:2008}
\bibinfo{author}{\bibfnamefont{M.}~\bibnamefont{Khodas}} \bibnamefont{and}
  \bibinfo{author}{\bibfnamefont{M.~G.} \bibnamefont{Vavilov}},
  \bibinfo{journal}{Phys. Rev. B} \textbf{\bibinfo{volume}{78}},
  \bibinfo{pages}{245319} (\bibinfo{year}{2008}).

\bibitem[{\citenamefont{Du et~al.}(2004)\citenamefont{Du, Zudov, Yang, Yuan,
  Pfeiffer, and West}}]{du:2004a}
\bibinfo{author}{\bibfnamefont{R.~R.} \bibnamefont{Du}},
  \bibinfo{author}{\bibfnamefont{M.~A.} \bibnamefont{Zudov}},
  \bibinfo{author}{\bibfnamefont{C.~L.} \bibnamefont{Yang}},
  \bibinfo{author}{\bibfnamefont{Z.~Q.} \bibnamefont{Yuan}},
  \bibinfo{author}{\bibfnamefont{L.~N.} \bibnamefont{Pfeiffer}},
  \bibnamefont{and} \bibinfo{author}{\bibfnamefont{K.~W.} \bibnamefont{West}},
  \bibinfo{journal}{Int. J. Mod. Phys. B} \textbf{\bibinfo{volume}{18}},
  \bibinfo{pages}{3465} (\bibinfo{year}{2004}).

\bibitem[{\citenamefont{Sassine et~al.}(2007)\citenamefont{Sassine, Krupko,
  Olshanetsky, Kvon, Portal, Hartmann, and Zhang}}]{sassine:2007}
\bibinfo{author}{\bibfnamefont{S.}~\bibnamefont{Sassine}},
  \bibinfo{author}{\bibfnamefont{Y.}~\bibnamefont{Krupko}},
  \bibinfo{author}{\bibfnamefont{E.}~\bibnamefont{Olshanetsky}},
  \bibinfo{author}{\bibfnamefont{Z.}~\bibnamefont{Kvon}},
  \bibinfo{author}{\bibfnamefont{J.}~\bibnamefont{Portal}},
  \bibinfo{author}{\bibfnamefont{J.}~\bibnamefont{Hartmann}}, \bibnamefont{and}
  \bibinfo{author}{\bibfnamefont{J.}~\bibnamefont{Zhang}},
  \bibinfo{journal}{Solid State Commun.} \textbf{\bibinfo{volume}{142}},
  \bibinfo{pages}{631} (\bibinfo{year}{2007}).

\bibitem[{\citenamefont{Kozlov et~al.}(2011)\citenamefont{Kozlov, Kvon,
  Mikhailov, Dvoretskii, and Portal}}]{kozlov:2011a}
\bibinfo{author}{\bibfnamefont{D.~A.} \bibnamefont{Kozlov}},
  \bibinfo{author}{\bibfnamefont{Z.~D.} \bibnamefont{Kvon}},
  \bibinfo{author}{\bibfnamefont{N.~N.} \bibnamefont{Mikhailov}},
  \bibinfo{author}{\bibfnamefont{S.~A.} \bibnamefont{Dvoretskii}},
  \bibnamefont{and} \bibinfo{author}{\bibfnamefont{J.~C.}
  \bibnamefont{Portal}}, \bibinfo{journal}{JETP Lett.}
  \textbf{\bibinfo{volume}{93}}, \bibinfo{pages}{1770} (\bibinfo{year}{2011}).

\bibitem[{\citenamefont{Vasiliadou et~al.}(1993)\citenamefont{Vasiliadou,
  M\"uller, Heitmann, Weiss, von Klitzing, Nickel, Schlapp, and
  L\"osch}}]{vasiliadou:1993}
\bibinfo{author}{\bibfnamefont{E.}~\bibnamefont{Vasiliadou}},
  \bibinfo{author}{\bibfnamefont{G.}~\bibnamefont{M\"uller}},
  \bibinfo{author}{\bibfnamefont{D.}~\bibnamefont{Heitmann}},
  \bibinfo{author}{\bibfnamefont{D.}~\bibnamefont{Weiss}},
  \bibinfo{author}{\bibfnamefont{K.}~\bibnamefont{von Klitzing}},
  \bibinfo{author}{\bibfnamefont{H.}~\bibnamefont{Nickel}},
  \bibinfo{author}{\bibfnamefont{W.}~\bibnamefont{Schlapp}}, \bibnamefont{and}
  \bibinfo{author}{\bibfnamefont{R.}~\bibnamefont{L\"osch}},
  \bibinfo{journal}{Phys. Rev. B} \textbf{\bibinfo{volume}{48}},
  \bibinfo{pages}{17145} (\bibinfo{year}{1993}).

\bibitem[{\citenamefont{Konstantinov and Kono}(2009)}]{konstantinov:2009}
\bibinfo{author}{\bibfnamefont{D.}~\bibnamefont{Konstantinov}}
  \bibnamefont{and} \bibinfo{author}{\bibfnamefont{K.}~\bibnamefont{Kono}},
  \bibinfo{journal}{Phys. Rev. Lett.} \textbf{\bibinfo{volume}{103}},
  \bibinfo{pages}{266808} (\bibinfo{year}{2009}).

\bibitem[{\citenamefont{Konstantinov and Kono}(2010)}]{konstantinov:2010}
\bibinfo{author}{\bibfnamefont{D.}~\bibnamefont{Konstantinov}}
  \bibnamefont{and} \bibinfo{author}{\bibfnamefont{K.}~\bibnamefont{Kono}},
  \bibinfo{journal}{Phys. Rev. Lett.} \textbf{\bibinfo{volume}{105}},
  \bibinfo{pages}{226801} (\bibinfo{year}{2010}).

\bibitem[{\citenamefont{Monarkha}(2011)}]{monarkha:2011}
\bibinfo{author}{\bibfnamefont{Y.~P.} \bibnamefont{Monarkha}},
  \bibinfo{journal}{Low Temp. Phys.} \textbf{\bibinfo{volume}{37}},
  \bibinfo{pages}{90} (\bibinfo{year}{2011}).

\bibitem[{\citenamefont{Studenikin et~al.}(2004)\citenamefont{Studenikin,
  Potemski, Coleridge, Sachrajda, and Wasilewski}}]{studenikin:2004}
\bibinfo{author}{\bibfnamefont{S.~A.} \bibnamefont{Studenikin}},
  \bibinfo{author}{\bibfnamefont{M.}~\bibnamefont{Potemski}},
  \bibinfo{author}{\bibfnamefont{P.~T.} \bibnamefont{Coleridge}},
  \bibinfo{author}{\bibfnamefont{A.~S.} \bibnamefont{Sachrajda}},
  \bibnamefont{and} \bibinfo{author}{\bibfnamefont{Z.~R.}
  \bibnamefont{Wasilewski}}, \bibinfo{journal}{Solid State Commun.}
  \textbf{\bibinfo{volume}{129}}, \bibinfo{pages}{341} (\bibinfo{year}{2004}).

\bibitem[{\citenamefont{Mani et~al.}(2004)\citenamefont{Mani, Narayanamurti,
  von Klitzing, Smet, Johnson, and Umansky}}]{mani:2004a}
\bibinfo{author}{\bibfnamefont{R.~G.} \bibnamefont{Mani}},
  \bibinfo{author}{\bibfnamefont{V.}~\bibnamefont{Narayanamurti}},
  \bibinfo{author}{\bibfnamefont{K.}~\bibnamefont{von Klitzing}},
  \bibinfo{author}{\bibfnamefont{J.~H.} \bibnamefont{Smet}},
  \bibinfo{author}{\bibfnamefont{W.~B.} \bibnamefont{Johnson}},
  \bibnamefont{and} \bibinfo{author}{\bibfnamefont{V.}~\bibnamefont{Umansky}},
  \bibinfo{journal}{Phys. Rev. B} \textbf{\bibinfo{volume}{70}},
  \bibinfo{pages}{155310} (\bibinfo{year}{2004}).

\bibitem[{\citenamefont{Hatke et~al.}(2011{\natexlab{a}})\citenamefont{Hatke,
  Khodas, Zudov, Pfeiffer, and West}}]{hatke:2011e}
\bibinfo{author}{\bibfnamefont{A.~T.} \bibnamefont{Hatke}},
  \bibinfo{author}{\bibfnamefont{M.}~\bibnamefont{Khodas}},
  \bibinfo{author}{\bibfnamefont{M.~A.} \bibnamefont{Zudov}},
  \bibinfo{author}{\bibfnamefont{L.~N.} \bibnamefont{Pfeiffer}},
  \bibnamefont{and} \bibinfo{author}{\bibfnamefont{K.~W.} \bibnamefont{West}},
  \bibinfo{journal}{Phys. Rev. B} \textbf{\bibinfo{volume}{84}},
  \bibinfo{pages}{241302(R)} (\bibinfo{year}{2011}{\natexlab{a}}).

\bibitem[{\citenamefont{Dobbie et~al.}(2012)\citenamefont{Dobbie, Myronov,
  Morris, Hassan, Prest, Shah, Parker, Whall, and Leadley}}]{dobbie:2012}
\bibinfo{author}{\bibfnamefont{A.}~\bibnamefont{Dobbie}},
  \bibinfo{author}{\bibfnamefont{M.}~\bibnamefont{Myronov}},
  \bibinfo{author}{\bibfnamefont{R.~J.~H.} \bibnamefont{Morris}},
  \bibinfo{author}{\bibfnamefont{A.~H.~A.} \bibnamefont{Hassan}},
  \bibinfo{author}{\bibfnamefont{M.~J.} \bibnamefont{Prest}},
  \bibinfo{author}{\bibfnamefont{V.~A.} \bibnamefont{Shah}},
  \bibinfo{author}{\bibfnamefont{E.~H.~C.} \bibnamefont{Parker}},
  \bibinfo{author}{\bibfnamefont{T.~E.} \bibnamefont{Whall}}, \bibnamefont{and}
  \bibinfo{author}{\bibfnamefont{D.~R.} \bibnamefont{Leadley}},
  \bibinfo{journal}{Appl. Phys. Lett.} \textbf{\bibinfo{volume}{101}},
  \bibinfo{eid}{172108} (\bibinfo{year}{2012}).

\bibitem[{not({\natexlab{b}})}]{note:10}
\bibinfo{note}{The overstretched Si$_{0.2}$Ge$_{0.8}$ buffer was grown on the
  inverse-graded SiGe virtual substrate on (001) Si. The concentration of
  background impurities in the Ge channel is estimated at $\lesssim 2 \times
  10^{14}$ cm$^{-3}$.}

\bibitem[{not({\natexlab{c}})}]{note:6}
\bibinfo{note}{Hall bars fabricated from the same wafer have shown a record
  high mobility of $\mu \approx 1.3 \cdot 10^6$
  cm$^2$/V$\cdot$s.\citep{dobbie:2012,hassan:2013}}.

\bibitem[{\citenamefont{Hassan et~al.}(2013)\citenamefont{Hassan, Mironov,
  Dobbie, Morris, Halpin, Shah, Myronov, Leadley, Gabani, Feher
  et~al.}}]{hassan:2013}
\bibinfo{author}{\bibfnamefont{A.~H.~A.} \bibnamefont{Hassan}},
  \bibinfo{author}{\bibfnamefont{O.~A.} \bibnamefont{Mironov}},
  \bibinfo{author}{\bibfnamefont{A.}~\bibnamefont{Dobbie}},
  \bibinfo{author}{\bibfnamefont{J.~H.} \bibnamefont{Morris}},
  \bibinfo{author}{\bibfnamefont{J.~E.} \bibnamefont{Halpin}},
  \bibinfo{author}{\bibfnamefont{V.~A.} \bibnamefont{Shah}},
  \bibinfo{author}{\bibfnamefont{M.}~\bibnamefont{Myronov}},
  \bibinfo{author}{\bibfnamefont{D.~R.} \bibnamefont{Leadley}},
  \bibinfo{author}{\bibfnamefont{S.}~\bibnamefont{Gabani}},
  \bibinfo{author}{\bibfnamefont{A.}~\bibnamefont{Feher}},
  \bibnamefont{et~al.}, \bibinfo{journal}{2013 IEEE Int. Sci. Conf. Electron.
  and Nanotech. (ELNANO) Proc.} p.~\bibinfo{pages}{51} (\bibinfo{year}{2013}).

\bibitem[{\citenamefont{Zhang et~al.}(2007{\natexlab{a}})\citenamefont{Zhang,
  Chiang, Zudov, Pfeiffer, and West}}]{zhang:2007a}
\bibinfo{author}{\bibfnamefont{W.}~\bibnamefont{Zhang}},
  \bibinfo{author}{\bibfnamefont{H.-S.} \bibnamefont{Chiang}},
  \bibinfo{author}{\bibfnamefont{M.~A.} \bibnamefont{Zudov}},
  \bibinfo{author}{\bibfnamefont{L.~N.} \bibnamefont{Pfeiffer}},
  \bibnamefont{and} \bibinfo{author}{\bibfnamefont{K.~W.} \bibnamefont{West}},
  \bibinfo{journal}{Phys. Rev. B} \textbf{\bibinfo{volume}{75}},
  \bibinfo{pages}{041304(R)} (\bibinfo{year}{2007}{\natexlab{a}}).

\bibitem[{\citenamefont{Zhang et~al.}(2007{\natexlab{b}})\citenamefont{Zhang,
  Zudov, Pfeiffer, and West}}]{zhang:2007c}
\bibinfo{author}{\bibfnamefont{W.}~\bibnamefont{Zhang}},
  \bibinfo{author}{\bibfnamefont{M.~A.} \bibnamefont{Zudov}},
  \bibinfo{author}{\bibfnamefont{L.~N.} \bibnamefont{Pfeiffer}},
  \bibnamefont{and} \bibinfo{author}{\bibfnamefont{K.~W.} \bibnamefont{West}},
  \bibinfo{journal}{Phys. Rev. Lett.} \textbf{\bibinfo{volume}{98}},
  \bibinfo{pages}{106804} (\bibinfo{year}{2007}{\natexlab{b}}).

\bibitem[{\citenamefont{Hatke et~al.}(2009{\natexlab{a}})\citenamefont{Hatke,
  Zudov, Pfeiffer, and West}}]{hatke:2009a}
\bibinfo{author}{\bibfnamefont{A.~T.} \bibnamefont{Hatke}},
  \bibinfo{author}{\bibfnamefont{M.~A.} \bibnamefont{Zudov}},
  \bibinfo{author}{\bibfnamefont{L.~N.} \bibnamefont{Pfeiffer}},
  \bibnamefont{and} \bibinfo{author}{\bibfnamefont{K.~W.} \bibnamefont{West}},
  \bibinfo{journal}{Phys. Rev. Lett.} \textbf{\bibinfo{volume}{102}},
  \bibinfo{pages}{066804} (\bibinfo{year}{2009}{\natexlab{a}}).

\bibitem[{\citenamefont{Hatke et~al.}(2011{\natexlab{b}})\citenamefont{Hatke,
  Zudov, Pfeiffer, and West}}]{hatke:2011b}
\bibinfo{author}{\bibfnamefont{A.~T.} \bibnamefont{Hatke}},
  \bibinfo{author}{\bibfnamefont{M.~A.} \bibnamefont{Zudov}},
  \bibinfo{author}{\bibfnamefont{L.~N.} \bibnamefont{Pfeiffer}},
  \bibnamefont{and} \bibinfo{author}{\bibfnamefont{K.~W.} \bibnamefont{West}},
  \bibinfo{journal}{Phys. Rev. B} \textbf{\bibinfo{volume}{83}},
  \bibinfo{pages}{121301(R)} (\bibinfo{year}{2011}{\natexlab{b}}).

\bibitem[{\citenamefont{Bogan et~al.}(2012)\citenamefont{Bogan, Hatke,
  Studenikin, Sachrajda, Zudov, Pfeiffer, and West}}]{bogan:2012}
\bibinfo{author}{\bibfnamefont{A.}~\bibnamefont{Bogan}},
  \bibinfo{author}{\bibfnamefont{A.~T.} \bibnamefont{Hatke}},
  \bibinfo{author}{\bibfnamefont{S.~A.} \bibnamefont{Studenikin}},
  \bibinfo{author}{\bibfnamefont{A.}~\bibnamefont{Sachrajda}},
  \bibinfo{author}{\bibfnamefont{M.~A.} \bibnamefont{Zudov}},
  \bibinfo{author}{\bibfnamefont{L.~N.} \bibnamefont{Pfeiffer}},
  \bibnamefont{and} \bibinfo{author}{\bibfnamefont{K.~W.} \bibnamefont{West}},
  \bibinfo{journal}{Phys. Rev. B} \textbf{\bibinfo{volume}{86}},
  \bibinfo{pages}{235305} (\bibinfo{year}{2012}).

\bibitem[{\citenamefont{Hatke et~al.}(2009{\natexlab{b}})\citenamefont{Hatke,
  Zudov, Pfeiffer, and West}}]{hatke:2009b}
\bibinfo{author}{\bibfnamefont{A.~T.} \bibnamefont{Hatke}},
  \bibinfo{author}{\bibfnamefont{M.~A.} \bibnamefont{Zudov}},
  \bibinfo{author}{\bibfnamefont{L.~N.} \bibnamefont{Pfeiffer}},
  \bibnamefont{and} \bibinfo{author}{\bibfnamefont{K.~W.} \bibnamefont{West}},
  \bibinfo{journal}{Phys. Rev. Lett.} \textbf{\bibinfo{volume}{102}},
  \bibinfo{pages}{086808} (\bibinfo{year}{2009}{\natexlab{b}}).

\bibitem[{\citenamefont{Hatke et~al.}(2009{\natexlab{c}})\citenamefont{Hatke,
  Zudov, Pfeiffer, and West}}]{hatke:2009c}
\bibinfo{author}{\bibfnamefont{A.~T.} \bibnamefont{Hatke}},
  \bibinfo{author}{\bibfnamefont{M.~A.} \bibnamefont{Zudov}},
  \bibinfo{author}{\bibfnamefont{L.~N.} \bibnamefont{Pfeiffer}},
  \bibnamefont{and} \bibinfo{author}{\bibfnamefont{K.~W.} \bibnamefont{West}},
  \bibinfo{journal}{Phys. Rev. B} \textbf{\bibinfo{volume}{79}},
  \bibinfo{pages}{161308(R)} (\bibinfo{year}{2009}{\natexlab{c}}).

\bibitem[{\citenamefont{Giuliani and Quinn}(1982)}]{giuliani:1982}
\bibinfo{author}{\bibfnamefont{G.~F.} \bibnamefont{Giuliani}} \bibnamefont{and}
  \bibinfo{author}{\bibfnamefont{J.~J.} \bibnamefont{Quinn}},
  \bibinfo{journal}{Phys. Rev. B} \textbf{\bibinfo{volume}{26}},
  \bibinfo{pages}{4421} (\bibinfo{year}{1982}).

\bibitem[{\citenamefont{Dmitriev et~al.}(2003)\citenamefont{Dmitriev, Mirlin,
  and Polyakov}}]{dmitriev:2003}
\bibinfo{author}{\bibfnamefont{I.~A.} \bibnamefont{Dmitriev}},
  \bibinfo{author}{\bibfnamefont{A.~D.} \bibnamefont{Mirlin}},
  \bibnamefont{and} \bibinfo{author}{\bibfnamefont{D.~G.}
  \bibnamefont{Polyakov}}, \bibinfo{journal}{Phys. Rev. Lett.}
  \textbf{\bibinfo{volume}{91}}, \bibinfo{pages}{226802}
  (\bibinfo{year}{2003}).

\bibitem[{not({\natexlab{d}})}]{note:8}
\bibinfo{note}{At $\eac=\eac_{2+}$, where $\lambda^2$ changes the most,
  $\lambda^2(T)/\lambda^2(0) \simeq \exp(-2\pi(k_B T)^2/\hbar\oc E_F)\approx
  0.7$ at $T = 4$ K.}

\bibitem[{not({\natexlab{e}})}]{note:4}
\bibinfo{note}{The rate of scattering on angle $\theta$ can be expressed in
  terms of angular harmonics, $\tau_n=\tau_{-n}$, as $\tau_\theta^{-1} =
  \sum\tau_n^{-1}e^{i n\theta}$. In this notation, $\tq^{-1}=\tau_0^{-1}$,
  $\ttr^{-1}=\tau_0^{-1}-\tau_1^{-1}$, and
  $\tst^{-1}=3\tau_0^{-1}-4\tau_1^{-1}+\tau_2^{-1}$.}

\bibitem[{not({\natexlab{f}})}]{note:1}
\bibinfo{note}{High-temperature deviation from $T^{-2}$-dependence of the MIRO
  amplitude at $\eac = \eac_{1+}$ can be attributed to $T$-dependence of $\tq$
  which was ignored in our analysis.}

\bibitem[{not({\natexlab{g}})}]{note:2}
\bibinfo{note}{Because of large lateral size of our 2DHS ($\approx 4$ mm),
  magnetoplasmon and cyclotron resonances should occur at nearly the same
  magnetic field.}

\bibitem[{\citenamefont{Hatke et~al.}(2012)\citenamefont{Hatke, Zudov, Watson,
  and Manfra}}]{hatke:2012b}
\bibinfo{author}{\bibfnamefont{A.~T.} \bibnamefont{Hatke}},
  \bibinfo{author}{\bibfnamefont{M.~A.} \bibnamefont{Zudov}},
  \bibinfo{author}{\bibfnamefont{J.~D.} \bibnamefont{Watson}},
  \bibnamefont{and} \bibinfo{author}{\bibfnamefont{M.~J.}
  \bibnamefont{Manfra}}, \bibinfo{journal}{Phys. Rev. B}
  \textbf{\bibinfo{volume}{85}}, \bibinfo{pages}{121306(R)}
  (\bibinfo{year}{2012}).

\bibitem[{\citenamefont{Hatke et~al.}(2013)\citenamefont{Hatke, Zudov, Watson,
  Manfra, Pfeiffer, and West}}]{hatke:2013}
\bibinfo{author}{\bibfnamefont{A.~T.} \bibnamefont{Hatke}},
  \bibinfo{author}{\bibfnamefont{M.~A.} \bibnamefont{Zudov}},
  \bibinfo{author}{\bibfnamefont{J.~D.} \bibnamefont{Watson}},
  \bibinfo{author}{\bibfnamefont{M.~J.} \bibnamefont{Manfra}},
  \bibinfo{author}{\bibfnamefont{L.~N.} \bibnamefont{Pfeiffer}},
  \bibnamefont{and} \bibinfo{author}{\bibfnamefont{K.~W.} \bibnamefont{West}},
  \bibinfo{journal}{Phys. Rev. B} \textbf{\bibinfo{volume}{87}},
  \bibinfo{pages}{161307(R)} (\bibinfo{year}{2013}).

\bibitem[{not({\natexlab{h}})}]{note:7}
\bibinfo{note}{We have used lower microwave frequency due to lack of calibrated
  attenuators for higher frequencies.}

\bibitem[{\citenamefont{Khodas et~al.}(2010)\citenamefont{Khodas, Chiang,
  Hatke, Zudov, Vavilov, Pfeiffer, and West}}]{khodas:2010}
\bibinfo{author}{\bibfnamefont{M.}~\bibnamefont{Khodas}},
  \bibinfo{author}{\bibfnamefont{H.~S.} \bibnamefont{Chiang}},
  \bibinfo{author}{\bibfnamefont{A.~T.} \bibnamefont{Hatke}},
  \bibinfo{author}{\bibfnamefont{M.~A.} \bibnamefont{Zudov}},
  \bibinfo{author}{\bibfnamefont{M.~G.} \bibnamefont{Vavilov}},
  \bibinfo{author}{\bibfnamefont{L.~N.} \bibnamefont{Pfeiffer}},
  \bibnamefont{and} \bibinfo{author}{\bibfnamefont{K.~W.} \bibnamefont{West}},
  \bibinfo{journal}{Phys. Rev. Lett.} \textbf{\bibinfo{volume}{104}},
  \bibinfo{pages}{206801} (\bibinfo{year}{2010}).

\bibitem[{\citenamefont{Mani et~al.}(2010)\citenamefont{Mani, Gerl, Schmult,
  Wegscheider, and Umansky}}]{mani:2010}
\bibinfo{author}{\bibfnamefont{R.~G.} \bibnamefont{Mani}},
  \bibinfo{author}{\bibfnamefont{C.}~\bibnamefont{Gerl}},
  \bibinfo{author}{\bibfnamefont{S.}~\bibnamefont{Schmult}},
  \bibinfo{author}{\bibfnamefont{W.}~\bibnamefont{Wegscheider}},
  \bibnamefont{and} \bibinfo{author}{\bibfnamefont{V.}~\bibnamefont{Umansky}},
  \bibinfo{journal}{Phys. Rev. B} \textbf{\bibinfo{volume}{81}},
  \bibinfo{pages}{125320} (\bibinfo{year}{2010}).

\bibitem[{not({\natexlab{i}})}]{note:9}
\bibinfo{note}{Analysis of Shubnikov-de Haas oscillations often yields a
  severely underestimated quantum lifetime due to macroscopic density
  fluctuations.}

\bibitem[{\citenamefont{Yang et~al.}(2002)\citenamefont{Yang, Zhang, Du,
  Simmons, and Reno}}]{yang:2002}
\bibinfo{author}{\bibfnamefont{C.~L.} \bibnamefont{Yang}},
  \bibinfo{author}{\bibfnamefont{J.}~\bibnamefont{Zhang}},
  \bibinfo{author}{\bibfnamefont{R.~R.} \bibnamefont{Du}},
  \bibinfo{author}{\bibfnamefont{J.~A.} \bibnamefont{Simmons}},
  \bibnamefont{and} \bibinfo{author}{\bibfnamefont{J.~L.} \bibnamefont{Reno}},
  \bibinfo{journal}{Phys. Rev. Lett.} \textbf{\bibinfo{volume}{89}},
  \bibinfo{pages}{076801} (\bibinfo{year}{2002}).

\bibitem[{\citenamefont{Pillarisetty}(2011)}]{pillarisetty:2011}
\bibinfo{author}{\bibfnamefont{R.}~\bibnamefont{Pillarisetty}},
  \bibinfo{journal}{Nature (London)} \textbf{\bibinfo{volume}{479}},
  \bibinfo{pages}{324} (\bibinfo{year}{2011}).

\bibitem[{\citenamefont{Bykov et~al.}(2006)\citenamefont{Bykov, Bakarov,
  Islamov, and Toropov}}]{bykov:2006}
\bibinfo{author}{\bibfnamefont{A.~A.} \bibnamefont{Bykov}},
  \bibinfo{author}{\bibfnamefont{A.~K.} \bibnamefont{Bakarov}},
  \bibinfo{author}{\bibfnamefont{D.~R.} \bibnamefont{Islamov}},
  \bibnamefont{and} \bibinfo{author}{\bibfnamefont{A.~I.}
  \bibnamefont{Toropov}}, \bibinfo{journal}{JETP Lett.}
  \textbf{\bibinfo{volume}{84}}, \bibinfo{pages}{391} (\bibinfo{year}{2006}).

\end{thebibliography}

\end{document}